\documentclass[aps,prd,10pt,amsmath,amssymb,reprint,groupedaddress, floatfix]{revtex4-2}
\usepackage[utf8]{inputenc}
\usepackage{graphicx}
\usepackage{latexsym}
\usepackage{epsfig}
\usepackage{amsfonts}
\usepackage{amsmath}
\usepackage{amssymb}
\usepackage{subfig}
\usepackage{float}
\usepackage{hyperref}
\hypersetup{colorlinks=true,linkcolor=blue,	filecolor=magenta, urlcolor= black}
\usepackage{mathtools}

\begin{document}


\title{Tidal love number of a static spherically symmetric anisotropic compact star}


\author{Shyam Das}\email{shyam\_das@associates.iucaa.in}
\affiliation{Department of Physics, P. D. Women's College, Jalpaiguri 735101, West Bengal, India.}

\author{Bikram Keshari Parida}\email{parida.bikram90.bkp@gmail.com}
\affiliation{Department of Physics, Pondicherry University, Kalapet, Puducherry 605014, India.}

\author{Ranjan Sharma}\email{rsharma@associates.iucaa.in}
\affiliation{Department of Physics, Cooch Behar Panchanan Barma University, Cooch Behar 736101, West Bengal, India.}

\date{\today}

\begin{abstract}

Tidal deformability of a coalescing neutron star subjected to an external tidal field plays an important role in our probe for the structure and properties of compact stars. In particular, the tidal love number provides valuable information about the external gravitational field responsible for deforming the star. In this article, we compute the tidal love number of a particular class of anisotropic stars and analyze the impacts of anisotropy and compactness on the tidal love number. 

\end{abstract}


\maketitle

\section{\label{sec1} Introduction}

Compact objects provide extreme conditions in terms of gravity and density and thus are unique astrophysical laboratories for studying general relativity and super nuclear-density matter. In general, compact objects exist in binaries comprising of either two neutron stars (NS-NS binaries) or a black hole (BH) and a neutron star (NS) (BH-NS binaries). Such binaries radiate away energy. The merger of these objects generates huge gravitational waves which has been experimentally verified in the recent past.

Compact stars provide perfect places for investigating the nature of particle interactions at very high densities in a natural way \cite{Nevermann2019}. Neutron stars (NS) are compact objects of very high energy density having approximate masses $1.5 M_{\odot}$ and radii $10^{5}$ times smaller than the Sun's radius. Therefore, they are perfect natural systems to study nuclear matter properties at high densities. In fact, density inside the core of a NS can be as high as several times the density that is reached inside a heavy atomic nuclei \cite{haensel_neutron_2007}. Despite attempts of many decades, we still lack a proper understanding of the thermodynamical behaviour inside a compact star.  The extreme conditions at the interior of a compact star comprising matter of uncertain composition have prompted many investigators to study its gross macroscopic properties within the framework of General Relativity. In order to understand the microscopic properties, the macroscopic properties such as NS masses and radii have been used as important tools to constrain its EOS. 

In this article, we explore the possibility of introducing tidal deformation as one of the astrophysically observable macroscopic properties which can be used to study the interior of a NS \cite{Chatziioannou2020a}. Tidal effects are finite-size effects arising on extended bodies when they are immersed in an external gravitational field. Like any other extended object, NS are tidally deformed under the influence of an external tidal field. The tidal deformability measures the star's quadrupole deformation in response to a companion perturbating star \cite{Hinderer_2010}. The induced quadrupole moment of the neutron star affects the binding energy of the system and increases the rate of emission of gravitational waves \cite{PhysRev.131.435, PhysRev.136.B1224, Blanchet_2014}. The tidal deformability has a very important role in the observation of coalescing NS with gravitational waves, and it has been used to probe the internal structure of NS. The Tidal Love Number characterizes how easy or difficult it would be to deform a NS away from sphericity \cite{Yagi_2013A}, \cite{Nevermann2019}. The tidal love number can be computed by following the standard methods available in the literature \cite{poisson_gravity:_2014, Cardoso_2017, Sennett_2017, Maselli_2018, Hinderer_2008w}.
	
Tidal properties of a NS has been observed to have a direct bearing on the emitted gravitational wave signal. The Advanced LIGO \cite{LIGO2015} and Advanced Virgo \cite{Acernese_2014} gravitational-wave detectors have made their first observation of a binary NS inspiral \cite{Abbott_2017}, an event known as GW170817. The LIGO observations lead to the experimental insight of the Tidal Love Number \cite{Abbott_2017}. Later, another signal emitted during a neutron star binary coalescence, known as GW190425, was detected. The latter signal was much weaker than GW170817 as it was originated from a much greater distance \cite{Abbott_2020}. Nevertheless, these observations has helped investigators to constraint many physical properties of NS such maximum masses and radii \cite{Margalit_2017,Bauswein_2017,Rezzolla_2018,Ruiz_2018,Annala_2018,Radice_2018,Most_2018,Tews_2018,De_2018,Abbott_2018,K_ppel_2019}.
	 
For ready references of the calculation on relativistic tidal love number (TLN), we refer to the following citations \cite{Hinderer_2008w, Damour_2009,  Binnington_2009, Landry_2015}. The algorithm has also been extended to slowly rotating extended compact objects as can be found in references \cite{pani2015tidal, Pani_2015, Landry_2015, Landry_2017, BOGUTA1977413, Abdelsalhin2019a}. As stated earlier, TLN defines the tidal deformability of the star present in an external tidal field such as the companion binary \cite{poisson_gravity:_2014}. In gravitational wave astronomy, it is an important parameter which affects the late-spiral GW from coalescence and thereby provide important information about the nature of the merging objects \cite{Flanagan_2008}. Most importantly, TLN can hold us to constrain the EOS of the NS \cite{Flanagan_2008, Abbott_2018}. Note that TLN of  a black hole is zero \cite{Binnington_2009, Damour_2009, Fang_2005, G_rlebeck_2015, Poisson_2015}. For relatively less compact objects, the dominant contribution to the tidal deformability comes from the `even-parity' quadrupole term $l$, which starts to impact the phase of the GW signal emitted in a binary at the fifth post-Newtonian (5PN) order \cite{Zhu2020a}. The leading order (6PN) term of even-parity tidal deformability has also been calculated \cite{Vines_2011}. The odd-parity  (or gravitomagnetic or mass-current) tidal deformability was calculated independently in ref~\cite{Damour_2009} \& \cite{Binnington_2009}. The choice of fluid properties also affects the odd-parity tidal deformability \cite{Landry_2015}, as shown in ref~\cite{Pani_2018}. The pioneer in this field was Yagi \cite{Yagi_2014} who for the first time estimated the impact of odd-parity tidal deformability on the gravitational waves phase evolution and then extended the work by analyzing the signal from GW170817 \cite{Jim_nez_Forteza_2018}.
	
In our work, we develop a method to estimate the TLN for a spherically symmetric and anisotropic relativistic star which is in static equilibrium. In a compact object, pressures may be different in radial, and transverse directions and the difference of radial pressure ($p_r$) and tangential pressure ($p_t$) is defined as pressure anisotropy. Incorporating anisotropy into the matter distribution of compact objects, several anisotropic models have been developed and investigated which include the works of Maurya and Gupta \cite{MG14}, Maurya {\em et al} \cite{MGD15}, Pandya {\em et al} \cite{ PTS15}, Murad \cite{M13}, Mafa Takisa {\em et al} \cite{ MM14a, MM14b}, Sunzu {\em et al} \cite{SM14a, SM14b}, Matondo and Maharaj \cite{KM16}, Karmakar {\em et al} \cite{260}, Abreu {\em et al} \cite{261}, Ivanov \cite{262}, Herrera {\em et al} \cite{263}, Mak and Harko \cite{264}, Sharma and Mukherjee \cite{265}, Harko and Mak \cite{266}, Herrera {\em et al} \cite{267}, Maharaj and Maartens \cite{268}, Gokhroo and Mehra \cite{269}, Chaisi and Maharaj \cite{270,271}, Thomas and co-workers \cite{275,277}, Das {\em et al} \cite{276}, Thirukkanesh and Maharaj \cite{279},  fully covarient framework by Raposo {\em et al} \cite{Raposo2019}, amongst others. Ruderman \cite{219} and Canuto \cite{220} have shown that anisotropy may develop inside highly dense compact stellar objects due to a variety of factors.  Kippenhahn and Weigert \cite{221} revealed that in relativistic stars anisotropy might occur due to the existence of a solid core or type $3A$ superfluid. Strong magnetic fields can also generate an anisotropic pressure inside a self-gravitating body \cite{Weber}. Anisotropy may also develop due to the slow rotation of fluids \cite{223}. A mixture of perfect and a null fluid may also be represented by an effective anisotropic fluid model \cite{224}. Local anisotropy may occur in astrophysical objects for various reasons such as viscosity, phase transition \cite{225}, pion condensation \cite{226} and the presence of strong electromagnetic field \cite{227}. The factors contributing to the pressure anisotropy have also been discussed by Dev and Gleiser \cite{228, 229} and Gleiser and Dev \cite{230}. Ivanov \cite{231} pointed out that influences of shear, electromagnetic field etc. on self-bound systems can be absorbed if the system is considered to be anisotropic. Self-bound systems composed of scalar fields, the so-called `boson stars' are naturally anisotropy \cite{232}. Same is true for wormholes \cite{233} and gravastars \cite{234, 235} as well. The shearing motion of the fluid can be considered as one of the reasons for the presence of anisotropy in a self-gravitating body \cite{236}.  Bowers and Liang \cite{237} have extensively discussed the underlying causes of pressure anisotropy in the stellar interior and analyzed the effects of anisotropic stress on the equilibrium configuration of relativistic stars. The point we would like to stress is that in the studies of relativistic compact stars, we find it worthwhile to consider anisotropic stress rather than an isotropic fluid distribution.

The paper has been organized as follows: In Section \ref{sec2}, significance of tidal love number has been discussed. Section \ref{sec3} deals with the finding a physically acceptable model which can be used to calculate the tidal love number. In the section \ref{sec4}, using the model, the  tidal love number $k_2$ for different NS has been estimated. $k_2$ has also been calculated for a given compactness $\mathcal{C}$ but different anisotropies $\alpha$. In Section \ref{sec5}, some concluding remarks have been made.

\section{\label{sec2} Tidal love number}

We consider a static spherically symmetric neutron star (NS) immersed in an external tidal field. In response to the tidal field, by developing a multipolar structure, the star will be deformed by the tidal force. This kind of situation occurs in coalescing binary systems where each component is tidally deformed by the gravitational field of its companion. The Tidal Love Number(TLN) characterizes the deformability of the NS away from sphericity \cite{PhysRevD.88.023009}. For mathematical simplicity, in our calculation, we shall restrict ourselves to quadrupole moments $\mathcal{Q}_{ij}$ only. This is reasonable if the two binary neutron stars remain sufficiently far away from each other. In such a situation, the quadrupole moment ($l = 2$) dominates over the multiple moments. $\mathcal{Q}_{ij}$ can be related with the external tidal field $\mathcal{E}_{ij}$ as \cite{Hinderer2008}
	\begin{align}
	\mathcal{Q}_{ij} = - \Lambda \mathcal{E}_{ij},\label{1}
	\end{align}
where, $\Lambda$ is the tidal deformability of the neutron star and it is related to the tidal love number $k_{2}$ as \cite{Hinderer2008}, 
	\begin{align}
	 k_{2} = \dfrac{3}{2}\Lambda \, R^{-5}.\label{2}
	\end{align}
The Tidal Love Number is dimensionless. The quadrupole fields $\mathcal{Q}_{ij}$ and $\mathcal{E}_{ij}$ can be expanded in tensor spherical harmonics $\mathcal{Y}_{ij}^{lm}$ as:
	\begin{align}
	\mathcal{E}_{ij} &= \sum_{m = -2 }^{2} \mathcal{E}_{m} \mathcal{Y}_{ij}^{2m} = \mathcal{E}_{0} \mathcal{Y}_{ij}^{20} = \mathcal{E} \mathcal{Y}_{ij}^{20} \, ,\\
	\mathcal{Q}_{ij} &= \sum_{m = -2 }^{2}  \mathcal{Q}_{m} \mathcal{Y}_{ij}^{2m} = \mathcal{Q}_{0} \mathcal{Y}_{ij}^{20} = \mathcal{Q} \mathcal{Y}_{ij}^{20}.
	\end{align}
	
In the second equality, the coordinate system was so oriented that the term became symmetric in $\phi$. The only component that is non-vanishing is the $m = 0$ component. We can rewrite equation \eqref{1} as, 
	
	\begin{align}
	\mathcal{Q} = - \Lambda \mathcal{E}. \label{5}
	\end{align}
	
Now the background metric  $^{(0)}g_{\mu \nu}(x^{\nu})$ corresponding to the neutron star, with a small perturbation $h_{\mu \nu}(x^{\nu})$ due to external tidal field, gets modified as,
	
	\begin{align}
	g_{\mu \nu}\left(x^{\nu}\right)=^{(0)} g_{\mu \nu}\left(x^{\nu}\right)+h_{\mu \nu}\left(x^{\nu}\right). \label{6}
	\end{align}
	
	We write the background geometry of the spherical static star in the standard form, 
	
	\begin{align}
	^{\left( 0\right)} ds^{2} &=^{\left( 0\right)} g_{\mu \nu }dx^{\mu }dx^{\nu } \nonumber\\ &=-e^{2 \nu(r) }dt^{2}+e^{2 \lambda(r) }dr^{2}+r^{2}\left( d\theta ^{2}+\sin ^{2}\theta d\phi ^{2}\right). \label{7}
	\end{align}
		
For the linearized metric perturbation $h_{\mu \nu }$, using the method as in Ref.~\cite{Regge1957} and \cite{Biswas2019}, we restrict ourselves to static $l = 2, \, m=0$ even parity perturbation. With these assumptions, the perturbed metric becomes,
	
	\begin{align}
	\resizebox{0.45\textwidth}{!}{$
	h_{\mu \nu}=\operatorname{diag}\left[H_{0}(r) e^{2 \nu}, H_{2}(r) e^{2 \lambda}, r^{2} K(r), r^{2} \sin ^{2} \theta K(r)\right] Y_{2 m}(\theta, \phi)$}. \label{8}
	\end{align}
	
For the spherically static metric \eqref{7}, the stress-energy tensor is given as \cite{Pretel2020a, Bowers1974, PhysRevD.88.084022, PhysRevD.85.124023},
	
	\begin{align}
	^{(0)}T_{\chi}^{\xi}=\left(\rho+p_{t}\right) u^{\xi} u_{\chi}+p_{t} g_{\chi}^{\xi}+\left(p_{r}-p_{t}\right) \eta^{\xi} \eta_{\chi},\label{9}
	\end{align}
	
with, $u^{\xi} u_{\xi} = -1$, $\eta^{\xi} \eta_{\xi} = 1$ \& $\eta^{\xi} u_{\xi} = 0 $.\\
	
Furthermore, the energy-momentum tensor is perturbed by a perturbation tensor $\delta T_{\chi}^{\xi}$ which is defined as, 
	
	\begin{align}
	T_{\chi}^{\xi} = ^{(0)}T_{\chi}^{\xi} + \delta T_{\chi}^{\xi}. \label{10}
	\end{align}
	
The non-zero components of $T_{\chi}^{\xi}$ are:
	
	\begin{align}
	T_{t}^{t} &= -   \dfrac{d\rho}{dp_{r}} \,\, \delta p_{r}\,\, Y(\theta ,\phi )-  \rho (r), \label{11}\\
	T_{r}^{r} &=     \delta p_{r}(r) Y(\theta ,\phi )+ p_{r}(r), \label{12}\\
	T_{\theta}^{\theta} &=   \frac{dp_{t}}{dp_{r}}  \delta p_{r}(r) Y(\theta ,\phi )+  p_{t}(r), \label{13}\\
	T_{\phi}^{\phi} &=    \frac{dp_{t}}{dp_{r}}  \delta p_{r}(r) Y(\theta ,\phi )+  p_{t}(r). \label{14}
	\end{align}
	
	With these perturbed quantities, we can write down the perturbed Einstein Field Equations as
	
	\begin{align}
	G_{\chi}^{\xi} = 8 \pi T_{\chi}^{\xi}. \label{15}
	\end{align}
	
	Where, the Einstein tensor $G_{\chi}^{\xi} $ is calculated using the metric $g_{\chi \xi} $.\\
	
	\subsection{Derivation of master equation and expression for tidal love number}
	
Using the background  field equations $^{(0)}G_{\chi}^{\xi} = 8 \pi ^{(0)}T_{\chi}^{\xi}$, we obtain the following results:
	
\begin{align}
	^{(0)}G_{t}^{t} &= 8 \pi ^{(0)}T_{t}^{t},\nonumber\\
	\Rightarrow \lambda'(r) &= \frac{8 \pi  r^2 e^{2 \lambda (r)} \rho (r)-e^{2 \lambda (r)}+1}{2 r}, \label{16}\\
	^{(0)}G_{r}^{r} &= 8 \pi ^{(0)}T_{r}^{r},\nonumber\\
	\Rightarrow \nu'(r) &= \frac{8 \pi  r^2 p_{r}(r) e^{2 \lambda (r)}+e^{2 \lambda (r)}-1}{2 r}. \label{17}
	\end{align}
	Note that $\nabla_{\xi}^{(0)}T_{\chi}^{\xi} = 0 $. Choosing $\xi = r$, we obtain,  

\begin{align}
	p_{r}'(r) = \frac{-r p_{r}(r) \nu '(r)-2 p_{r}(r)+2 p_{t}(r)-r \rho (r) \nu '(r)}{r}. \label{18}
	\end{align}
	
	For the perturbed metric, using Einstein equations  \eqref{15}, we get the following results,
		
\begin{align}
	&G_{\theta}^{\theta} - G_{\phi}^{\phi} = 0 \Rightarrow H_{0}(r) = H_{2}(r) = H(r), \label{19}\\
	&G_{r}^{\theta} =0 \Rightarrow K' = H' +2 H \nu' , \label{20}\\
	&G_{\theta}^{\theta} + G_{\phi}^{\phi} = 8 \pi (T_{\theta}^{\theta} + T_{\phi}^{\phi}), \nonumber \\
	&\Rightarrow \delta p_{r} = \frac{H(r) e^{-2 \lambda (r)} \left(\lambda '(r)+\nu '(r)\right)}{8 \pi  \frac{dp_{t}}{dp_{r}} r}. \label{21}
	\end{align}
	
Now, using the identity,
\begin{align*}
    \resizebox{0.45\textwidth}{!}{$\dfrac{\partial^{2} Y(\theta,\phi)}{\partial \theta^{2}} + cot(\theta) \dfrac{\partial Y(\theta,\phi)}{\partial \theta}+ \csc ^2(\theta ) \dfrac{\partial^{2} Y(\theta,\phi)}{\partial \phi^{2}} = -6 Y(\theta ,\phi )$},
\end{align*}
 eqn \eqref{16}, \eqref{17}, \eqref{18}, \eqref{19}, \eqref{20} \& \eqref{21}, we obtain the master equation for $H(r)$ as,
		
\begin{align}
	&- \frac{1}{e^{-2 \lambda (r)} Y(\theta ,\phi )} \left[ G_{t}^{t} - G_{r}^{r}\right] = - \frac{8 \pi}{e^{-2 \lambda (r)} Y(\theta ,\phi )} \left[ T_{t}^{t} - T_{r}^{r}\right]\nonumber\\
	&\Rightarrow H''(r) + \mathcal{R} H'(r) + \mathcal{S} H(r) =0, \label{22}
	\end{align}
Where, 
	\begin{align}
	    \mathcal{R} = - \left(\frac{-e^{2 \lambda (r)}-1}{r}-4 \pi  r e^{2 \lambda (r)} (p_{r}(r)-\rho (r))\right), \label{23}
	\end{align}
	
\begin{widetext}
	\begin{align}
	    \mathcal{S} &= - \left(\frac{4 e^{2 \lambda (r)}+e^{4 \lambda (r)}+1}{r^2}+64 \pi ^2 r^2 p_{r}(r)^2 e^{4 \lambda (r)}+ 16 \pi  e^{2 \lambda (r)} \left(p_{r}(r) \left(e^{2 \lambda (r)}-2\right)-p_{t}(r)-\rho (r)\right)+\nonumber \right.\\
	&\left. \frac{-4 \pi  \frac{d\rho}{dp_{r}} e^{2 \lambda (r)} (p_{r}(r)+\rho (r))-4 \pi  e^{2 \lambda (r)} (p_{r}(r)+\rho (r))}{\frac{dp_{t}}{dp_{r}}}\right). \label{24}
	\end{align}
	\end{widetext}
	
The exterior region  of the static spherically symmetric star will be described by Schwarzschild metric and hence by setting, $\rho = 0,\, p_{r} = 0, \, p_t=0$ and $e^{2 \lambda} = 1/(1- 2 M/r)$, the master equation \eqref{22} takes the form,
	
	\begin{align}
	\resizebox{0.47\textwidth}{!}{$
	    -H''(r)-\frac{2 (M-r) H'(r)}{r (2 M-r)}+\frac{2 H(r) \left(2 M^2-6 M r+3 r^2\right)}{r^2 (r-2 M)^2} =0$}. \label{25}
	\end{align}
	
	The solution to this second-order differential equation \eqref{25} is obtained as,

	\begin{widetext}
\begin{align}
H(r) =&\frac{c_2 \left(-2 M \left(2 M^3+4 M^2 r-9 M r^2+3 r^3\right)-3 r^2 (r-2 M)^2 \log \left(\frac{r}{M}-2\right)+3 r^2 (r-2 M)^2 \log \left(\frac{r}{M}\right)\right)}{2 M^2 r (2 M-r)} \nonumber\\
&+  \frac{3 c_1 r (2 M-r)}{M^2}. \label{26}
	\end{align} 
	\end{widetext}
	
Where, $c_1$ and $c_2$ are integration constants. In order to get the expression for these constants, let us make a series expansion of the equation \eqref{26},
	
\begin{align}
H(r) = -\frac{3 c_1 r^2}{M^2}+\frac{6 c_1 r}{M}-\frac{c_2 \left(8 M^3\right)}{5 r^3}+\mathcal{O}\left(\left(\frac{1}{r}\right)^4\right). \label{27}
\end{align}
	
Now in the star's local asymptotic rest frame, at large $r$ the metric coefficient $g_{tt}$ is given by \cite{Thorne1998, Hinderer2008, PhysRevD.34.3617},
	
	\begin{align}
	\frac{\left(1-g_{t t}\right)}{2}=&-\frac{M}{r}-\frac{3 \mathcal{Q}_{i j}}{2 r^{3}}\left(n^{i} n^{j}-\frac{1}{3} \delta^{i j}\right)+\mathcal{O}(\frac{1}{r^{3}} ) \nonumber \\
	&+\frac{1}{2} \mathcal{E}_{i j} x^{i} x^{j}+ \mathcal{O}(r^{3}), \label{28}
	\end{align}
Where, $n^{i} = x^{i}/r$. Matching the asymptotic solution using equation \eqref{27} together with the expansion of equation \eqref{28} and using equation \eqref{1}, we have,
	
\begin{align}
	    c_1 = - \frac{M^2 \mathcal{E}}{3}, \quad c_2 = \frac{15 \mathcal{Q}}{8 M^3}. \label{29}
	\end{align}
	
Finally, the expression for tidal love number $k_2$ can be obtained by using equations \eqref{2}, \eqref{29} and \eqref{26} and also using the expression for $H(r)$ and its derivatives at the star's surface $r=R$ as,
	
	\begin{align}
	    k_2 = [8 (1-2 \mathcal{C})^2 \mathcal{C}^5 (2 \mathcal{C} (\mathit{y}-1)-\mathit{y}+2)]/ X,\label{30}
	\end{align}
	Where,
	\begin{widetext}
	\begin{align}
X &= (5(2 \mathcal{C} (\mathcal{C} (2 \mathcal{C} (\mathcal{C} (2 \mathcal{C} (\mathit{y}+1)+3 \mathit{y}-2)-11 \mathit{y}+13)+3 (5 \mathit{y}-8))-3 \mathit{y}+6) \nonumber\\
&\left.\left.+3 (1-2 \mathcal{C})^2 (2 \mathcal{C} (\mathit{y}-1)-\mathit{y}+2) \log \left(\frac{1}{\mathcal{C}}-2\right)-3 (1-2 \mathcal{C})^2 (2 \mathcal{C} (\mathit{y}-1)-\mathit{y}+2) \log \left(\frac{1}{\mathcal{C}}\right)\right)\right), \label{31}
	\end{align}
	\end{widetext}
	
	Note that $\mathcal{C} = \frac{M}{R}$ and $ \mathit{y}$ depend on $r,\, H(r)$ and it's derivatives evaluated at $R$ in the form,
	
	\begin{align}
	\mathit{y} =\dfrac{r H'(r)}{H(r)}_R. \label{32}
	\end{align}
	
To calculate the tidal love number $k_2$ for a particular compact star, we need to specify a model which we can be utilized to calculate $y$ and subsequently $k_2$ for a particular NS of given mass $M$ and radius $R$.
	
\section{\label{sec3} CHOOSING A PHYSICALLY ACCEPTABLE MODEL}
\subsection{Einstein field equations:}
To describe the interior of a static and spherically symmetric relativistic star, we write the line element in coordinates $(x^{a}) = (t,r,\theta,\phi)$ as 
\begin{equation}
 ds^{2} = -e^{2\nu(r)} dt^{2} + e^{2\lambda(r)} dr^{2} +
 r^{2}(d\theta^{2} + \sin^{2}{\theta} d\phi^{2}). \label{33}
\end{equation}

We also assume an anisotropic matter distribution for which the energy-momentum tensor is assumed in the form, 
\begin{equation}
 T_{ij}=\mbox{diag}(-\rho,~ p_r,~ p_t,~ p_t).\label{34}
\end{equation}

The energy density $\rho$, the radial pressure $p_r$ and  the tangential pressure $p_t$ are measured relative to the comoving fluid velocity $u^i = e^{-\nu}\delta^i_0.$  For the line element \eqref{33}, the independent set of Einstein field equations are then obtained as,
\begin{eqnarray}
 \rho &=& \frac{1}{r^{2}} \left[ r(1-e^{-2\lambda})
\right]',\label{35}\\
 p_r &=& - \frac{1}{r^{2}} \left( 1-e^{-2\lambda} \right)
+
\frac{2\nu'}{r}e^{-2\lambda} ,\label{36}\\
p_t &=& e^{-2\lambda}\left( \nu'' + \nu'^{2} +
\frac{\nu'}{r}- \nu'\lambda' - \frac{\lambda'}{r} \right) ,\label{37}
\end{eqnarray}
where primes ($'$) denote differentiation with respect to $r$. In the field equations \eqref{35}-\eqref{37}, we have assumed $8\pi G=1=c$. The system of equations determines the behaviour of the gravitational field of an anisotropic imperfect fluid sphere. The mass contained within a radius $r$ of the sphere is defined as,
\begin{equation}
 m(r)= \frac{1}{2}\int_0^r\omega^2 \rho(\omega)d\omega. \label{38}
\end{equation}

We define, $\Delta = p_t-p_r$ as the measure of anisotropy. The anisotropic stress will be directed outward (repulsive) when $p_t > p_r$ (i.e., $ \Delta > 0$) and inwards when $p_t < p_r$ (i.e.,  $\Delta < 0$). 

\subsection{ A particular  anisotropic model}

To calculate the tidal love number, we choose a particular model which is an anisotropic generalization of the Korkina and Orlyanskii solution III  obtained earlier by \cite{ThiruSharma}. To examine the physical acceptability of the solution, we first write the variables which are obtained as,
 \begin{eqnarray}
e^{2\nu} &=&A^2(1+aCr^2)^2, \label{39}\\
 e^{2\lambda} &=& \Big[1-BCr^2(1+3aCr^2)^{-2/3}-\alpha Cr^2 (1+aCr^2)^{-1} \nonumber \\ && \times (1+3aCr^2)^{-2/3}  \Big]^{-1}, \label{40}\\
\Delta &=&\frac{\alpha C^2 a r^2(1+3aC r^2)^{1/3}}{(1+a C r^2)^3}. \label{41}
\end{eqnarray}

The line element \eqref{33} then takes the form,
\begin{eqnarray}
\label{eq:a18} ds^{2} &=& -A^2(1+aCr^2)^2 dt^{2}
\nonumber\\
&& + \Big[1-B C r^2(1+3 a C r^2)^{-2/3}   -\alpha C r^2
(1+a C r^2)^{-1} \nonumber \\ && \times (1+3 a C r^2)^{-2/3}  \Big]^{-1}dr^{2}  + r^{2}(d\theta^{2} + \sin^{2}{\theta} d\phi^{2}).
\end{eqnarray}
For an isotropic sphere ($\alpha=0$), if we set $B=0$ and $C=1$, the metric \eqref{eq:a18} reduces to, 
\begin{equation}
\label{eq:a19} ds^{2} = -A^2(1+ar^2)^2 dt^{2}  + dr^{2} +
r^{2}(d\theta^{2} + \sin^{2}{\theta} d\phi^{2}).
\end{equation}
which is the Korkina and Orlyanskii solution III\cite{Korkina}. In other words,  the solution \eqref{eq:a18} obtained by \cite{ThiruSharma} is an anisotropic generalization of the solution of Korkina and Orlyanskii\cite{Korkina}.

\subsection{Physical acceptability of the solution}

We now examine the physical acceptability of our solution: 

\begin{enumerate}
\item[i.]
In this model, we have $(e^{2\nu(r)})'_{r=0}=(e^{2\lambda(r)})'_{r=0}=0 $
and  $e^{2\nu (0)}=A^2,~ e^{2\lambda(0)}=1$; these imply that the metric is regular at the centre $r=0$.

\item[ii.]  Since $\rho(0)= 3\,C(B+\alpha)$ and $\displaystyle p_r(0)=p_t(0)=C (4a-B- \alpha)$, the energy density, radial pressure and tangential pressure will be non-negative at the centre if we choose the parameters satisfying the condition $a>\frac{B+\alpha}{4}$.

\item[iii.] The interior solution \eqref{33} should be matched to the exterior  Schwarzschild metric
\begin{eqnarray}
 \label{eq:exterior}
ds^2 &=& - \left( 1 - \frac{2M}{r} \right) dt^2 +\left( 1 -
\frac{2M}{r} \right)^{-1} dr^2 \nonumber\\
&&+ r^2 (d\theta^2 +\sin^2\theta d\phi^2), 
\end{eqnarray}
across the boundary boundary of the star $r=R$, where $M$ is the total mass of the sphere which can be obtained directly from Eq.\eqref{38} as
$$\displaystyle M=m(s)=\frac{C R^3(B+a B C R^2+\alpha)}{2(1+2 a C R^2)(1+3 a C R^2)^{\frac{2}{3}}}.$$
Matching of the line elements \eqref{33} and (\ref{eq:exterior}) at the boundary $r=R$ yields,
\begin{eqnarray}
 \label{eq:f43}\left( 1 - \frac{2M}{R}
\right)&=&1-\frac{\alpha CR^2}{(1+aCR^2)(1+3 a C R^2)^{\frac{2}{3}}} \nonumber \\ && -\frac{B C R^2}{(1+3 a C R^2)^{\frac{2}{3}}},\\
\label{eq:f44}\left( 1 - \frac{2M}{R} \right)&=&A^2(1+a C R^2)^2,
\end{eqnarray}
Making use of the junction conditions, the constants $A, \, B, \, C$ are determined as

\begin{eqnarray}
A&=&\frac{(5 M -2 R)}{2\sqrt{R(R-2 M)}},\label{52}\\
C&=&\frac{M}{a R^2(2 R-5 M)},\label{53}\\
B&=&-\frac{(5 M-2R)[2^{\frac{8}{3}}a(\frac{M-R}{5 M -2R})^{\frac{2}{3}}(2 M -R)+R \alpha]}{2R(2 M-R)}.\label{54}
\end{eqnarray}

\item[iv.] The gradient of density, radial pressure and tangential pressure are respectively obtained as,

\begin{widetext}
\begin{eqnarray}
\frac{d\rho}{dr}&=& \frac{(2 a C^2 r (-10 B (1 + a C r^2)^4 + (-15 + a C r^2 (-53 + a C r^2 (-49 + 5 a C r^2))) \alpha))}{((1 + 
   a C r^2)^3 (1 + 3 a C r^2)^{\frac{8}{3}})},\\
\frac{dp_r}{dr}&=&-\frac{1}{((1 + a C r^2)^3 (1 +3 a C r^2)^{\frac{5}{3}}))}\times (((2 a C^2 r (-2 B (1 + a C r^2) (-1 + 5 a^2 C^2 r^4) + a (4 (1 +  3 a C r^2)^{\frac{2}{3}}  \nonumber\\
 && + C r^2 (a (16 (1 + 3 a C r^2)^{2/3} + C r^2 (12 a (1 + 3 a C r^2)^{2/3} - 25 \alpha)) - 8 \alpha)) + \alpha))),\\
\frac{dp_t}{dr}&=&\frac{1}{(1 + a C r^2)^4 (1 + 3 a C r^2)^{\frac{5}{3}}}\times (4 a C^2 r (B (1 + a C r^2)^2 (-1 + 5 a^2 C^2 r^4) +  a (-2 (1 + 3 a C r^2)^{\frac{2}{3}}  \nonumber\\
 && + C r^2 (6 \alpha + a (-10 (1 + 3 a C r^2)^{\frac{2}{3}} + C r^2 (17 \alpha + a (-14 (1 + 3 a C r^2)^{\frac{2}{3}}  \nonumber\\
 &&  + C r^2 (-6 a (1 + 3 a C r^2)^{\frac{2}{3}} + 5 \alpha)))))))).
\end{eqnarray}
\end{widetext}

The decreasing nature of these quantities is shown graphically.

\item[v.] Within a stellar interior, it is expected that the speed of sound should be less than the speed of light i.e., $\displaystyle 0
\leq \frac{dp_r}{d\rho}\leq 1$ and $\displaystyle 0 \leq
\frac{dp_t}{d\rho}\leq 1$.\\
In this model, we have,
\begin{widetext}
\begin{eqnarray}
\frac{dp_r}{d\rho}&=&-\frac{1}{(-10 B (1 + a C r^2)^4 + (-15 +  a C r^2 (-53 + a C r^2 (-49 + 5 a C r^2))) \alpha))}\times \Big[(((1 + 3 a C r^2)   \nonumber\\
&& (-2 B (1 + a C r^2) (-1 + 5 a^2 C^2 r^4) + a (4 (1 + 3 a C r^2)^{\frac{2}{3}} + C r^2 (a (16 (1 + 3 a C r^2)^{\frac{2}{3}} + \nonumber\\
&&  C r^2 (12 a (1 + 3 a C r^2)^{\frac{2}{3}} - 25 \alpha)) - 8 \alpha)) + \alpha)) \Big],\\
\frac{dp_t}{d\rho}&=&\frac{1}{( 10 B (1 + a C r^2)^5 + 15 \alpha +  a C r^2 (68 + a C r^2 (102 + a C r^2 (44 - 5 a C r^2))) \alpha))} \times \Big[ -((2 (1 + 3 a C r^2) \nonumber\\
&& (B (1 + a C r^2)^2 (-1 + 5 a^2 C^2 r^4) +  a (-2 (1 + 3 a C r^2)^{\frac{2}{3}} +  C r^2 (6 \alpha + a (-10 (1 + 3 a C r^2)^{\frac{2}{3}} \nonumber\\
&& + C r^2 (17 \alpha +  a (-14 (1 + 3 a C r^2)^{\frac{2}{3}} + C r^2 (-6 a (1 + 3 a C r^2)^{\frac{2}{3}} + 5 \alpha)))))))) \Big].
\end{eqnarray}
\end{widetext}

By choosing the model parameters appropriately, it can be shown that this requirement is also satisfied in this model.

\item[vi.] Fulfilment of the energy conditions for an anisotropic fluid i.e.,  $\rho +p_r+2p_t\geq 0$, $\rho +p_r \geq 0$ \& $\rho +p_t \geq 0$ can also be shown graphically to be satisfied in this model.
\end{enumerate}

\subsection{Physical behaviour of the model}

Physical quantities in this model are obtained as,

\begin{widetext}
\begin{eqnarray}
\rho &=& \frac{(B C (1 + a C r^2)^2 (3 + 5 a C r^2) + 
 C (3 + a C r^2 (6 - a C r^2)) \alpha)}{((1 + a C r^2)^2 (1 + 
   3 a C r^2)}^{\frac{5}{3}},\label{44}\\
p_r &=& -\frac{((C (B (1 + a C r^2) (1 + 5 a C r^2) + \alpha + 
       a (-4 (1 + 3 a C r^2)^{\frac{2}{3}} + 
          C r^2 (-4 a (1 + 3 a C r^2)^{\frac{2}{3}} + 5 \alpha))))}{((1 + 
       a C r^2)^2 (1 + 3 a C r^2)^{\frac{2}{3}}))},\label{45}\\
p_t &=& \dfrac{1}{((1 + a C r^2)^3 (1 +3 a C r^2)^{\frac{2}{3}})} \times \Big[(C (-B (1 + a C r^2)^2 (1 + 5 a C r^2) + a (4 (1 + 3 a C r^2)^{\frac{2}{3}} + C r^2 (2 a (4 (1 + 3 a C r^2)^{\frac{2}{3}} \nonumber\\ && + 
            C r^2 (2 a (1 + 3 a C r^2)^{\frac{2}{3}} - \alpha)) - 
         5 \alpha)) - \alpha)) \Big],\label{46} \\
\Delta &=& \frac{(a C^2 r^2 (1 + 3 a C r^2)^{\frac{1}{3}} \alpha)}{(1 + a C r^2)^3},\label{47}\\
m(r)&=& \frac{C r^3 (B + a B C r^2 + \alpha)}{2 (1 + a C r^2) (1 + 3 a C r^2)^{\frac{2}{3}}}. \label{48}
\end{eqnarray}
\end{widetext}

\begin{figure}[H]
\includegraphics[width=0.44\textwidth]{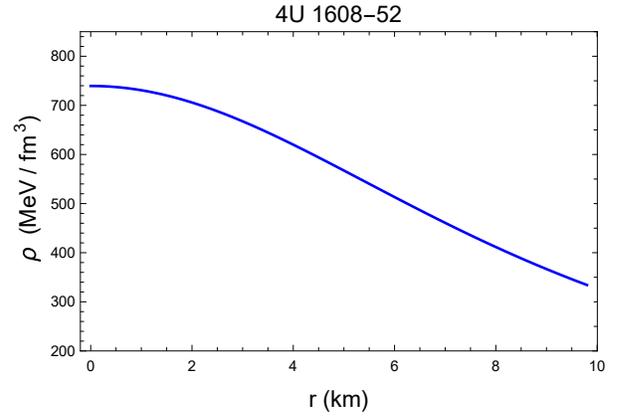}
\caption{Density profiles.}
\label{fig1}
\end{figure}

\begin{figure}[htbp]
\includegraphics[width=0.45\textwidth]{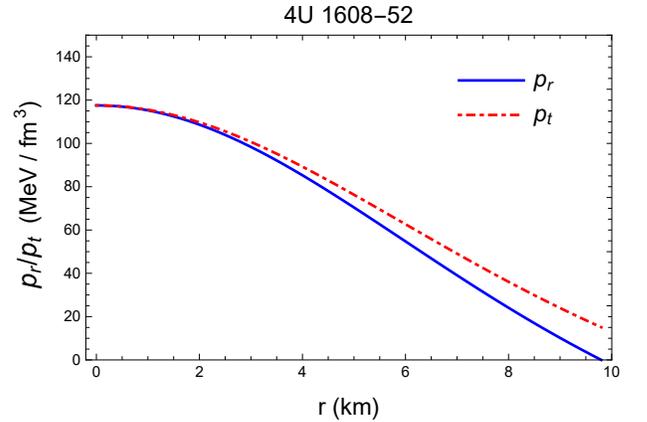}
\caption{Radial and tangential pressure profiles.}
\label{fig2}
\end{figure}

\begin{figure}[htbp]
\includegraphics[width=0.45\textwidth]{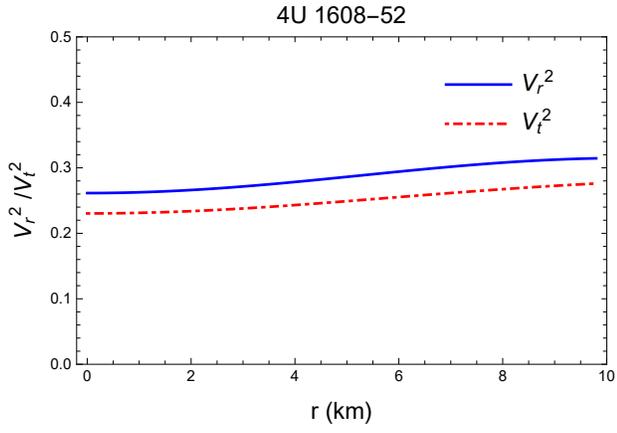}
\caption{Radial \& transverse component of sound speed.}
\label{fig3}
\end{figure}

\begin{figure}[htbp]
\includegraphics[width=0.45\textwidth]{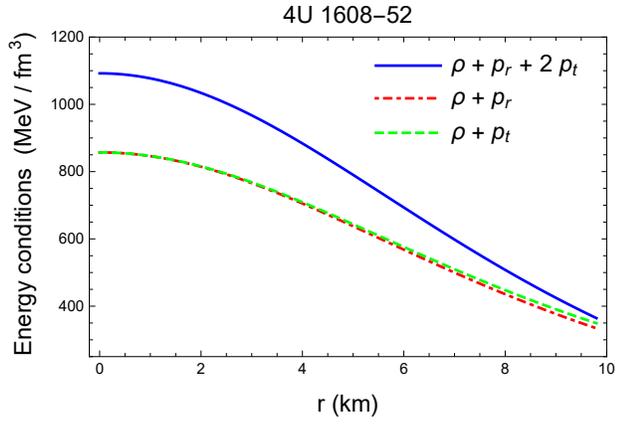}
\caption{Fulfillment of energy condition.}
\label{fig4}
\end{figure}

\begin{figure}[htbp]
\includegraphics[width=0.45\textwidth]{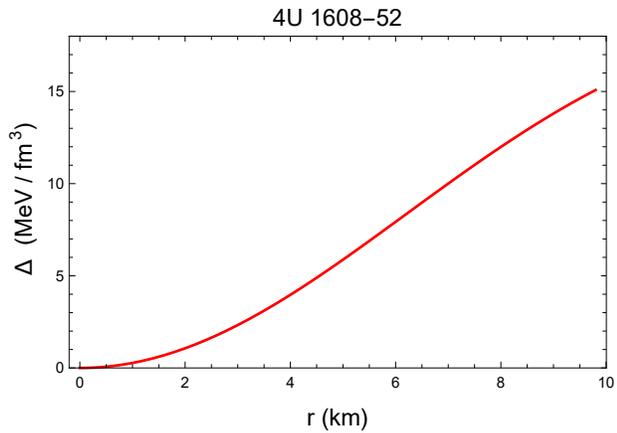}
\caption{Radial variation of anisotropy.}
\label{fig5}
\end{figure}

\begin{figure}[htbp]
\includegraphics[width=0.45\textwidth]{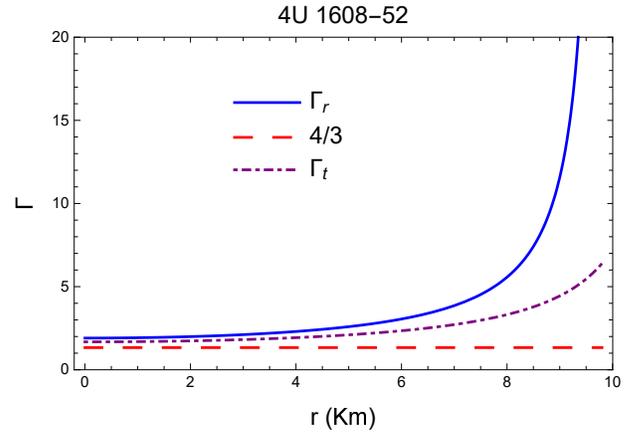}
\caption{Radial variation of adiabatic index.}
\label{fig6}
\end{figure}

\begin{figure}[H]
\includegraphics[width=0.45\textwidth]{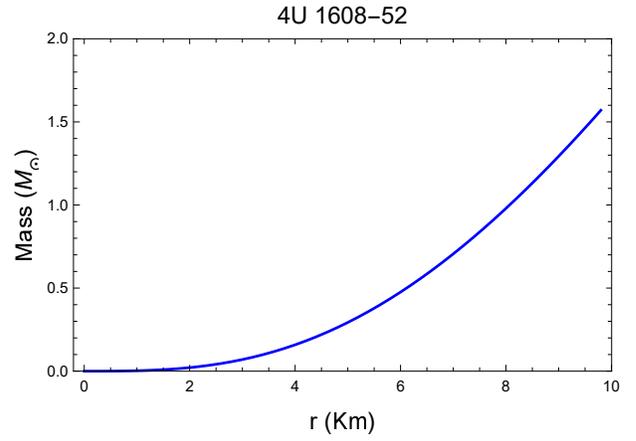}
\caption{Radial variation of mass.}
\label{fig7}
\end{figure}

\begin{figure}[H]
\includegraphics[width=0.45\textwidth]{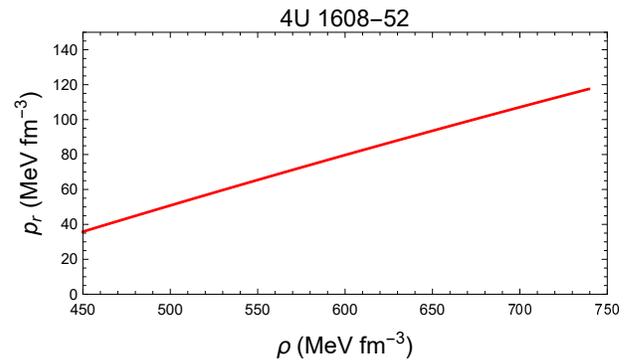}
\caption{Equation of state.}
\label{fig8}
\end{figure}

The simple elementary functional forms of the physical quantities help us to make a detailed study of the physical behaviour of the star. Most importantly, the solution contains an `anisotropic switch' $\alpha$, which allows us to investigate the impact of anisotropy effectively. We analyze the physical behaviour of the model by using the values of masses and radii of observed pulsars as input parameters.  We consider the data available from the pulsar $ 4U 1608-52$ whose estimated mass and radius are $M = 1.57 ~M_{\odot}$ and $R = 9.8~km$, respectively \cite{Roupas2020}. For these values, we determine two sets of constants. For an isotropic case ($\alpha=0$),  we obtain the constants $B = 2.48125,~C = 0.00300606,~A=0.56353$ and assuming the star to be composed of an anisotropic fluid distribution (we assume $\alpha =1$), the constants are calculated as $A=0.56353,~B = 1.70527,~C = 0.0030606$. Note that the parameter $a$ remains free in this model which we set as $a=1$ in both the cases. Making use of these values, we show graphically the nature of all the physically meaningful quantities in Fig.~(\ref{fig1})-(\ref{fig8}). The plots clearly show that all the quantities comply with the requirements of a realistic star. In particular, the figures highlight the effect of anisotropy on the gross physical behaviour of the compact star.

\begin{figure}[H]
\includegraphics[width=0.45\textwidth]{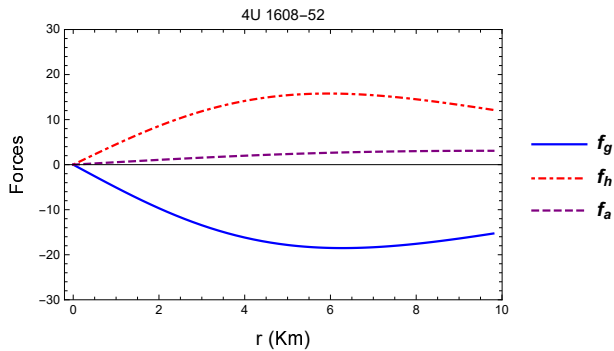}
\caption{Forces in equilibrium.}
\label{fig9}
\end{figure}

\section{\label{sec4}Numerical calculation of tidal love number}

Using the method employed in \cite{Rahmansyah2020}, we can calculate the numerical value of $k_2$ for a particular neutron star. To this end, we rewrite the master equation \eqref{22} using the equation \eqref{32} as,

\begin{align}
    r y' +y^2 + (r \mathcal{R} - 1) y + r^2  \mathcal{S} = 0. \label{b60}
\end{align}
	
From equation \eqref{30}, at $\mathcal{C} = 0$ we expect $k_2 = 0$. This implies that $y(0) = 2$. Moreover, at the horizon formation limit $\mathcal{C} = 0.5$, the tidal love number $k_2$ vanishes for all values of $y$.
	
\begin{figure}[htbp]
	    \centering
	    \includegraphics[width=0.45\textwidth]{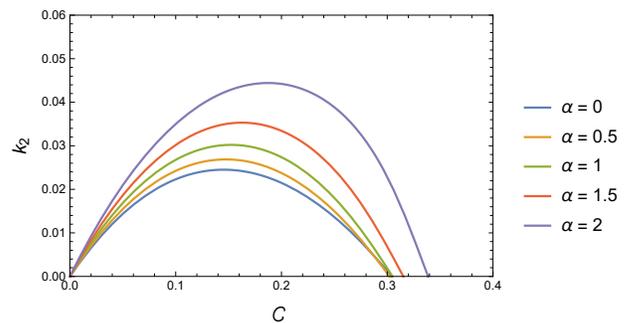}
	    \caption{$k_2$ is plotted against $\mathcal{C} $ for different $\alpha$.}
	    \label{fig:a10}
	\end{figure}

In order to solve the differential equation \eqref{b60}, we use the initial condition $y(0) = 2$ in addition to the expression for $\mathcal{R}, \, \mathcal{S}$  using equations \eqref{23} \& \eqref{24}, respectively. Using the initial condition and equations \eqref{40}, \eqref{44}, \eqref{45} and \eqref{46} for a particular NS, equation \eqref{b60} can be solved and subsequently using equation \eqref{30}, the tidal love number $k_2$ can be calculated. One can also find the analytical expression for $y(r)$ in terms of compactness factor $\mathcal{C}$ and anisotropy $\alpha$. Employing this technique, we plot the relation between $k_2$ and compactness factor $\mathcal{C}$ for different values of $\alpha$ as shown in the figure \ref{fig:a10}. We note that $k_2$ increases gradually with increasing $\mathcal{C}$ up to a certain value and then decreases with further increase of $\mathcal{C}$. For a  NS having compactness $\mathcal{C} < 3.4$ \& $\alpha \leq 2$, the above scheme can be used to calculate the tidal love number in this model.  
	
\begin{figure}[htbp]
	    \includegraphics[scale =0.45]{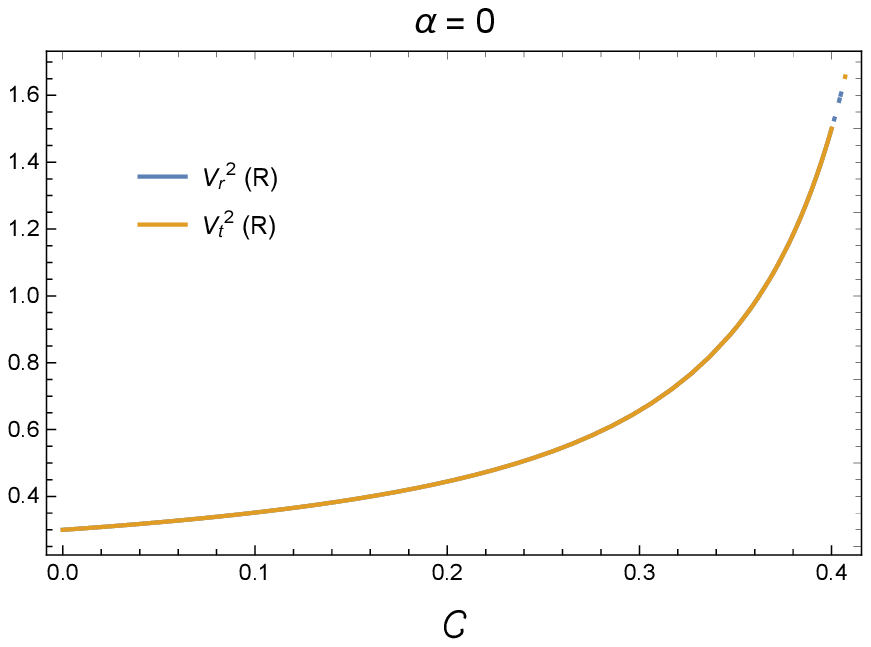}
	    \includegraphics[scale =0.45]{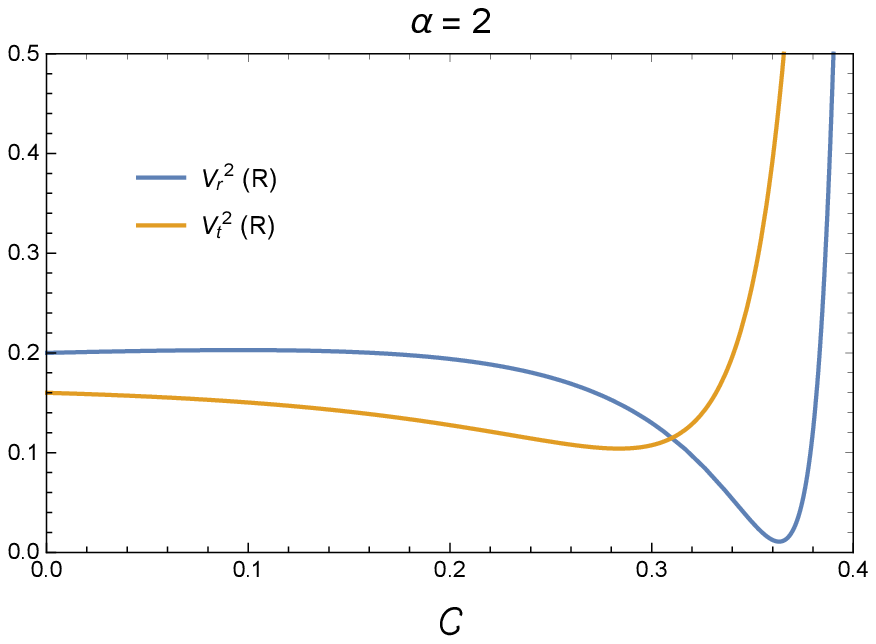}
	     \includegraphics[scale =0.45]{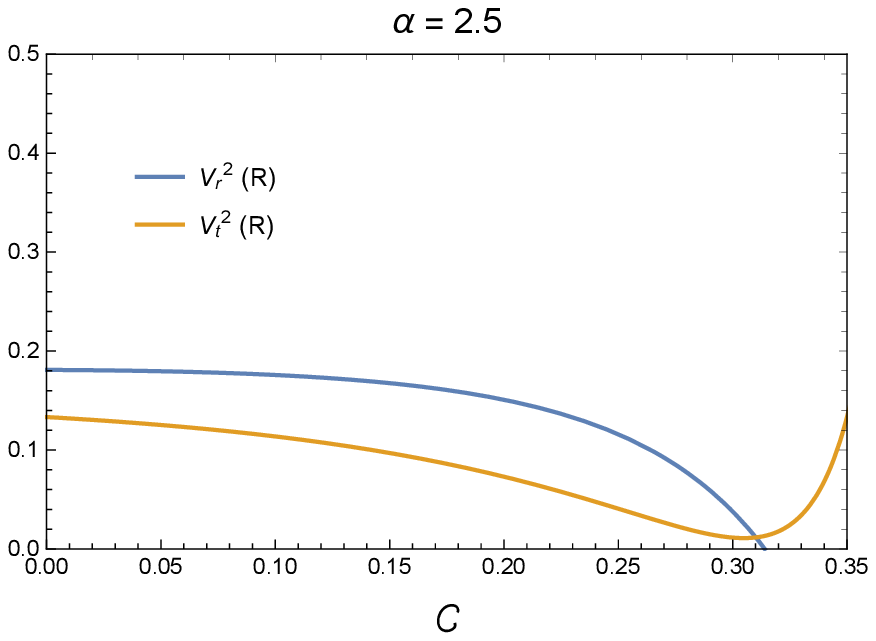}
	      \includegraphics[scale =0.45]{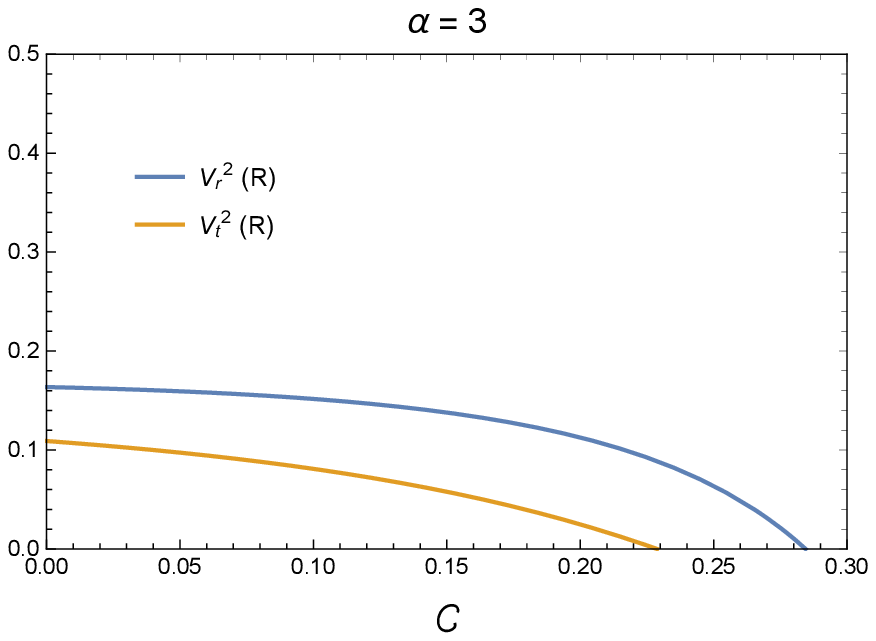}
	       \includegraphics[scale =0.45]{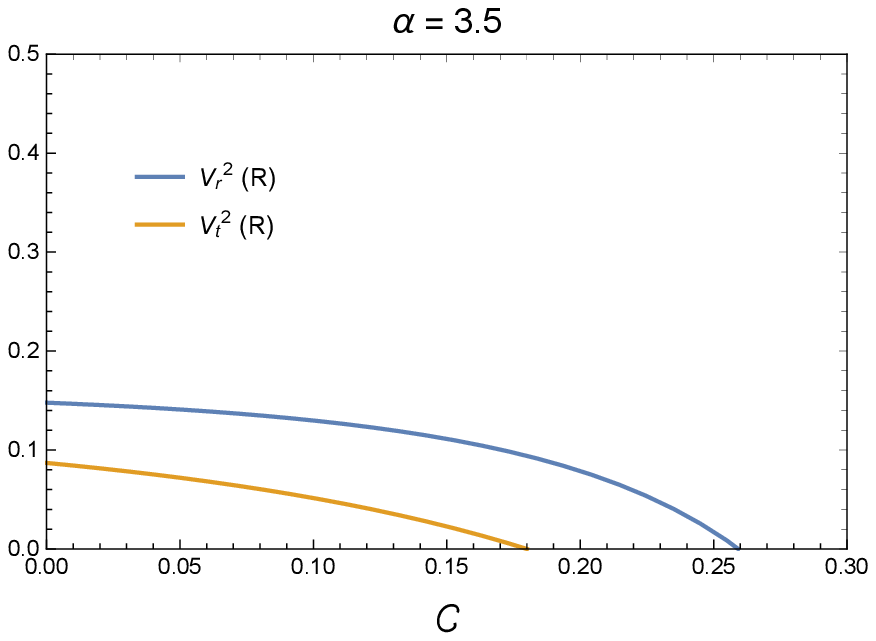}
	        \includegraphics[scale =0.45]{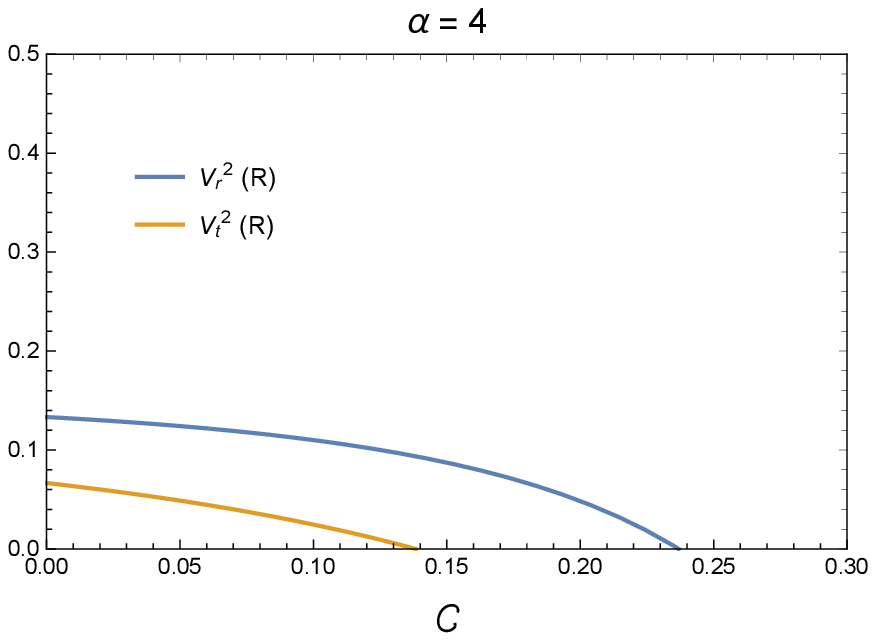}
	    \caption{$V_r ^2 (R) $ \& $V_t ^2 (R)$ are plotted against $\mathcal{C}$ for different $\alpha$.}
	    \label{fig:a11}
	\end{figure}

For $\alpha > 2$, a discontinuity arises in the plot of $k_2$ vs $\mathcal{C}$. To address this problem, we set a maximum limit on $\mathcal{C}$ for a particular $\alpha > 2$, using the `physical acceptability' conditions discussed earlier. One can check that for all range of values of $\mathcal{C} <0.34$, $\rho(r=0), \rho(r= R), p_r(r = 0), p_r(r = R), p_t(r = 0), p_t(r =  R) \geq 0$. The energy condition $\rho + p_r + 2 p_t \geq 0$. Therefore, the maximum limit on $\mathcal{C}$ can be calculated from the condition $0 \leq \dfrac{dp_r}{d \rho} \leq 1$ and $0 \leq \dfrac{d p_t}{d \rho} \leq 1$. For different values of $\alpha$, $V_r ^2(R) = \dfrac{dp_r}{d \rho}$ and $V_t ^2(R) = \dfrac{d p_t}{d \rho} $ are plotted against $\mathcal{C}$  in figure \ref{fig:a11}. It then becomes easy to evaluate the maximum limit on $\mathcal{C}$ for different $\alpha$ from the plot. For example, in figure \ref{fig:a11}, we note that for $0 \leq \alpha \leq 2$, the range of compactness is $0 \leq \mathcal{C} \leq 0.4$. For $0 \leq \alpha \leq 2$, figure \ref{fig:a10} shows that $0 \leq \mathcal{C} \leq 0.34$. In table \ref{tab:a1}, the maximum value of $\mathcal{C}$ for different $\alpha$ is given.

\begin{figure}[htbp]
	    \includegraphics[scale = 0.45]{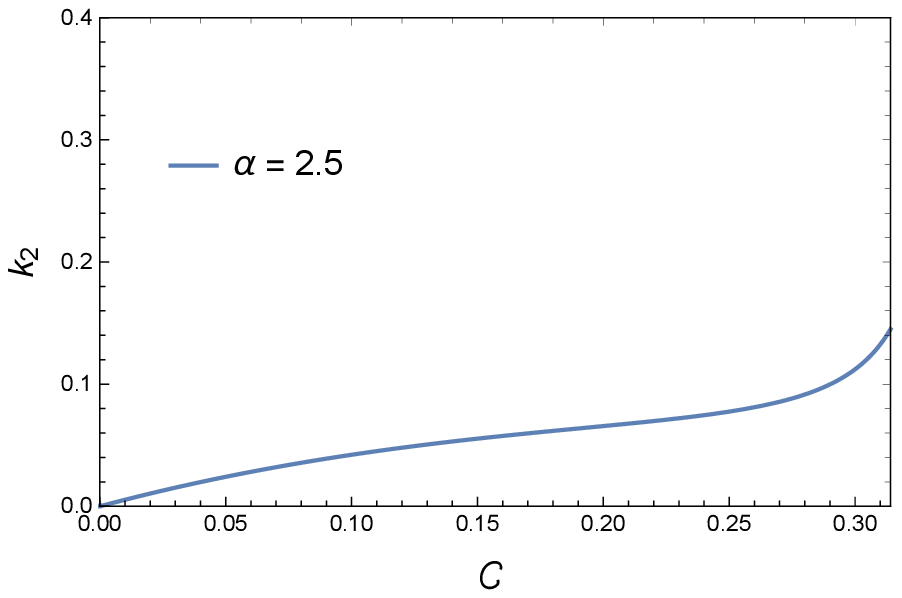}
	    \includegraphics[scale = 0.45]{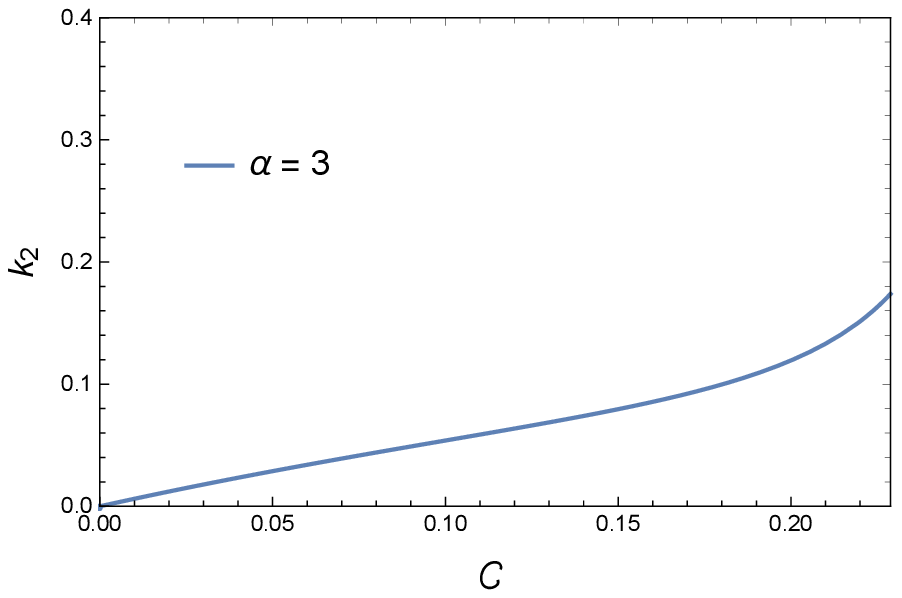}
	    \includegraphics[scale = 0.45]{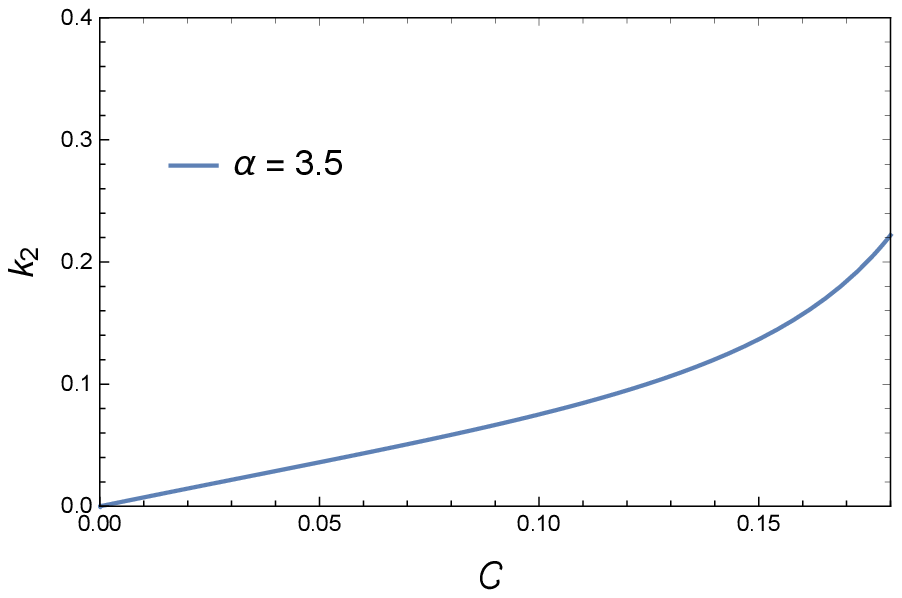}
	    \includegraphics[scale = 0.45]{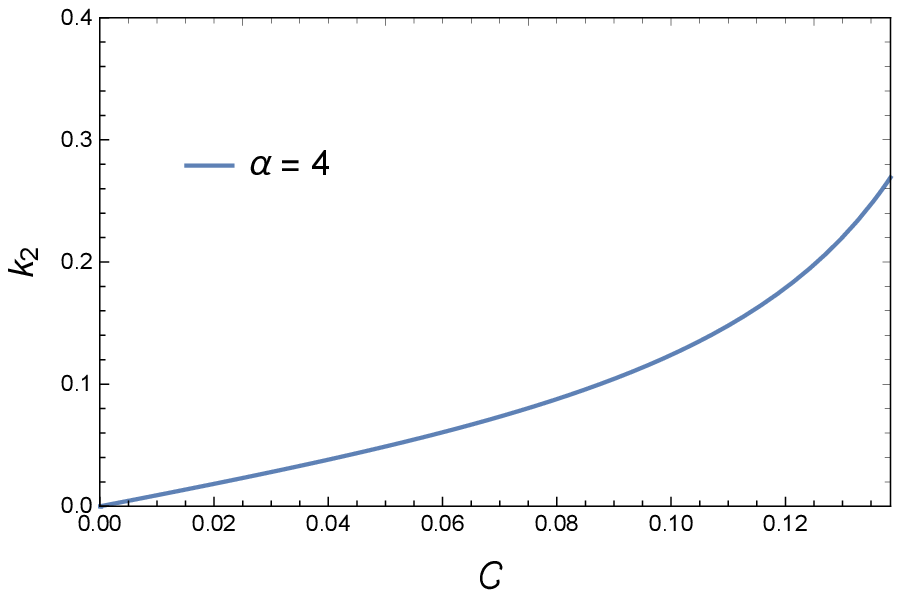}
	    \caption{$k_2$ is plotted against $\mathcal{C}$ for different $\alpha$. Only physically allowed range of $\alpha$ \& $\mathcal{C}$ are considered.}
	    \label{fig:a12}
	\end{figure}

	For $\alpha > 2 $, variation of the tidal love number $k_2$ against $\mathcal{C}$ is shown in figure \ref{fig:a12}. In table \ref{tab:a1}, the numerical values of the tidal love number $k_2$ is shown for different $\alpha$.
	
	\begin{table}[htbp]
	\caption{The maximum limit of $\mathcal{C}$ and its corresponding $k_2$ for $\alpha >2 $.}
	    \begin{tabular}{|c|c|c|}
	    \hline
	    Anisotropy $\alpha$ & Compactness  $\mathcal{C}$ & Tidal love number $k_2$\\
	    \hline \hline
	    
	      2.5   &  0.31405 & 0.144954 \\
	       3  &  0.2288 & 0.173846\\
	       3.5 & 0.1800 & 0.221919\\
	       4 & 0.1384 & 0.269166\\
	       \hline
	    \end{tabular}
	    \label{tab:a1}
	\end{table}

	\begin{figure}[htbp]
	    \centering
	    \includegraphics[scale = 0.75]{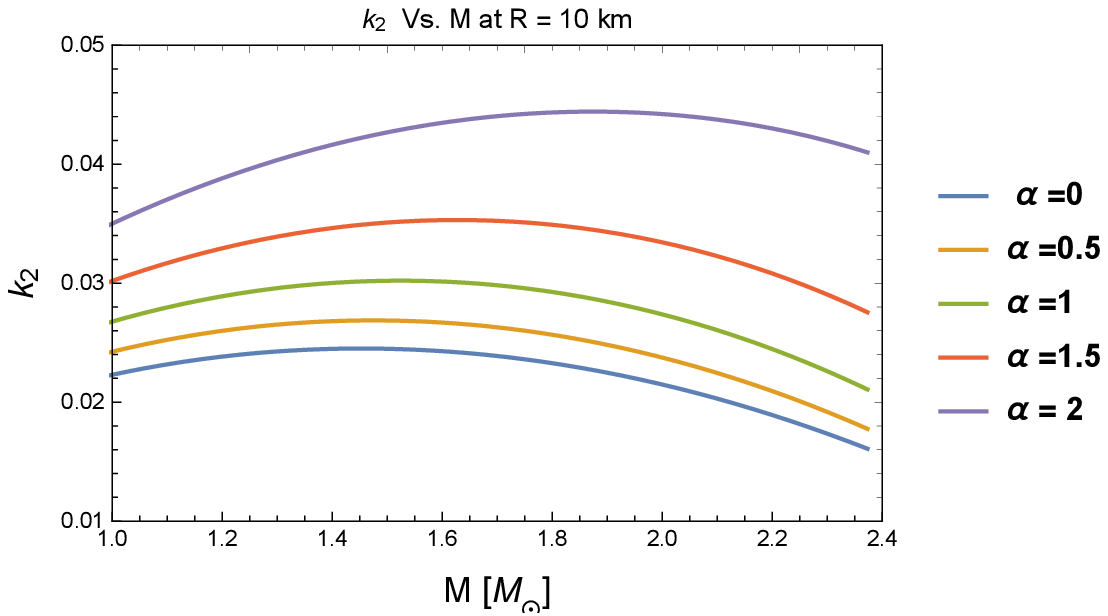}
	    \includegraphics[scale = 0.75]{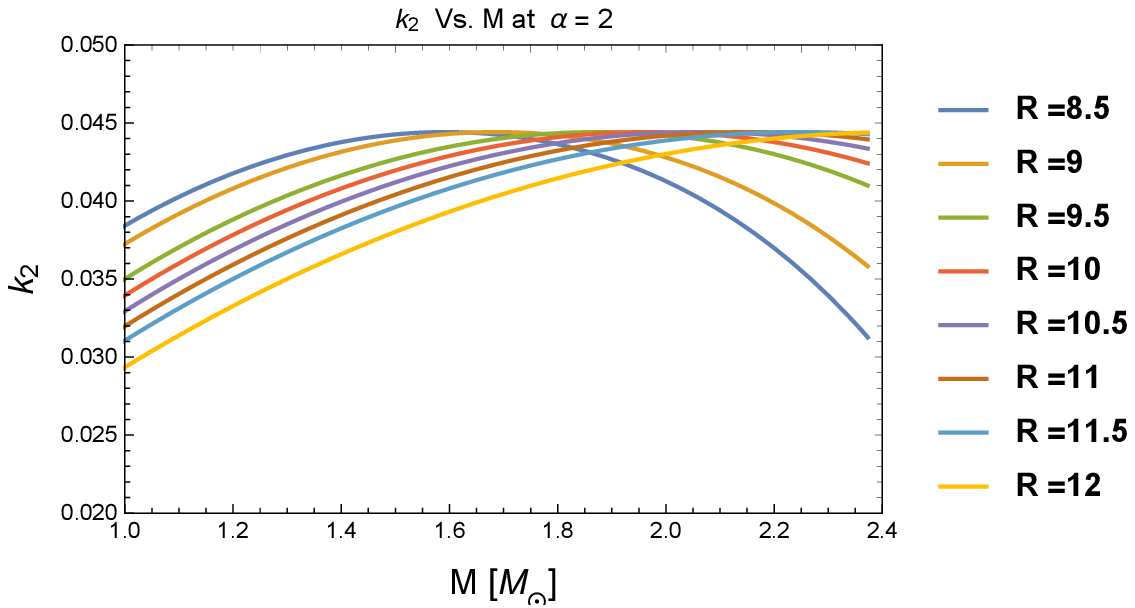}
	    \caption{(Top) $k_2$ is plotted against mass  $M$ for different $\alpha$ at $R = 10\, km$, (Bottom) $k_2$ is plotted against mass $M$ for different radius $R$ at $\alpha = 2$.}
	    \label{fig:x13}
	\end{figure}

	\begin{figure}[htbp]
	    \centering
	    \includegraphics[scale =0.74]{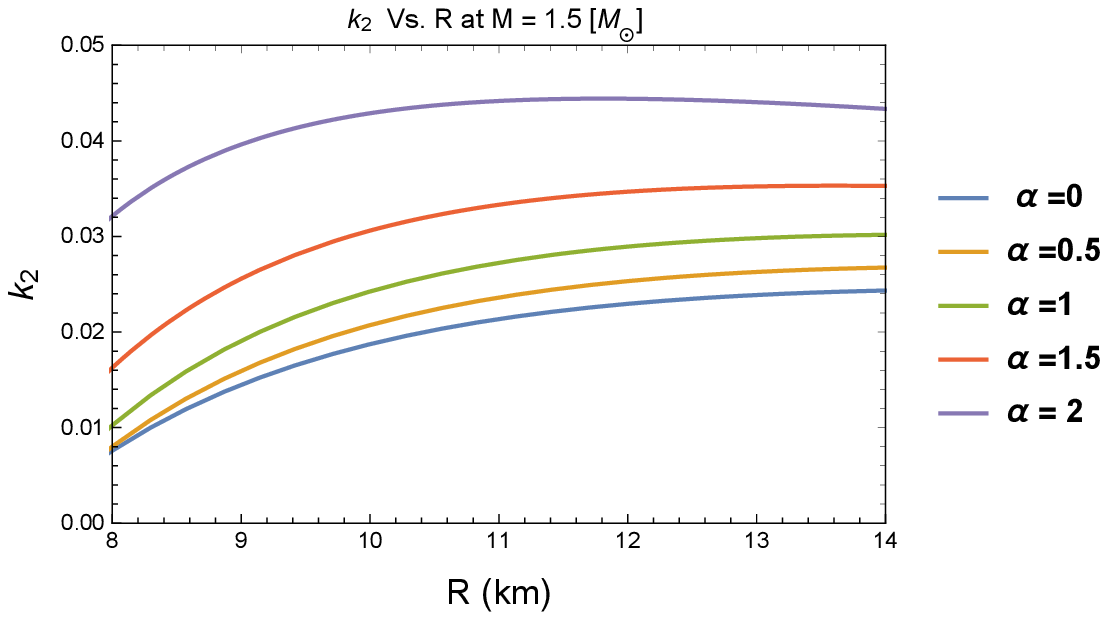}
	    \includegraphics[scale =0.74]{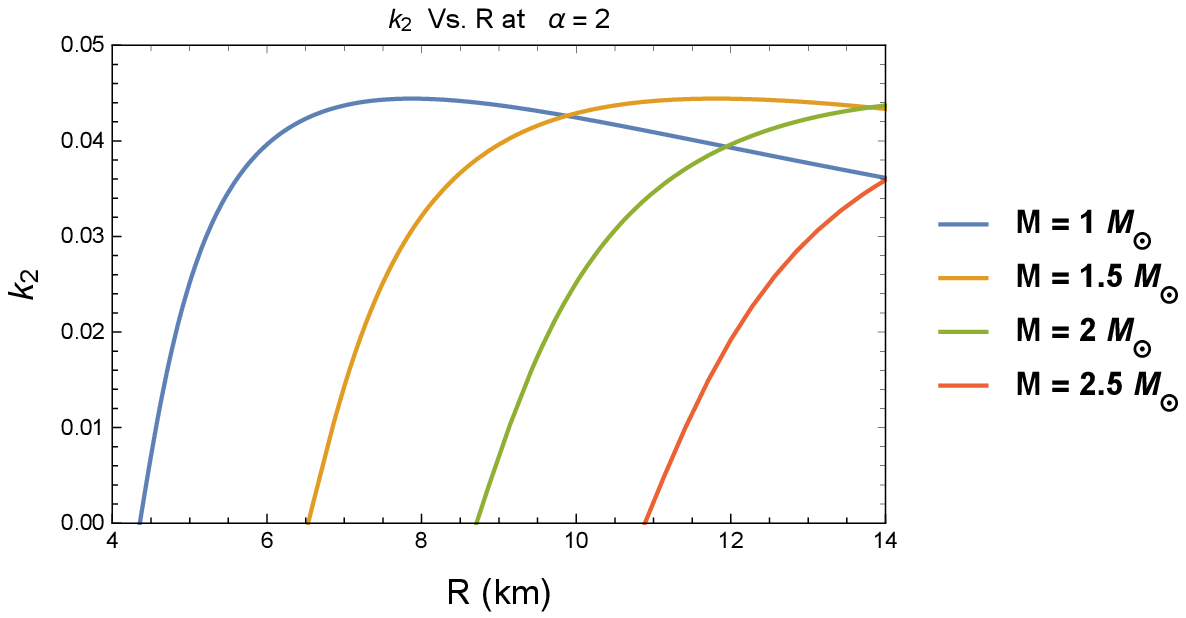}
	    \caption{(Top) $k_2$ is plotted against radius $R$ for different $\alpha$ at mass $M = 1.5 \, M_{\odot}$, (Bottom) $k_2$ is plotted against $R$ for different mass $M$ at $\alpha = 2$.}
	    \label{fig:x14}
	\end{figure}
	
	In the fig.\ref{fig:x13}, (Top) $k_2$ is plotted against mass $M$ for different $\alpha$ at $R = 10\, km$, (Bottom) $k_2$ is plotted against mass $M$ for different radius $R$ at $\alpha = 2$. In the fig.\ref{fig:x14}, (Top) $k_2$ is plotted against radius $R$ for different $\alpha$ at mass $M = 1.5 \, M_{\odot}$ and (Bottom) $k_2$ is plotted against $R$ for different mass $M$ at $\alpha = 2$.
	
	\begin{figure}[htbp]
	    \includegraphics[scale=0.65]{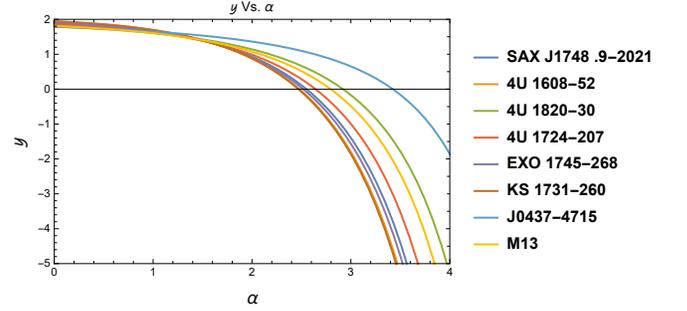}
	    \caption{ $y$ is plotted against $\alpha$ for different compact objects. Only physically allowed range of $\alpha$ is considered.}
	    \label{fig:b10}
	\end{figure}

	\begin{figure}[htbp]
	    \includegraphics[scale=0.65]{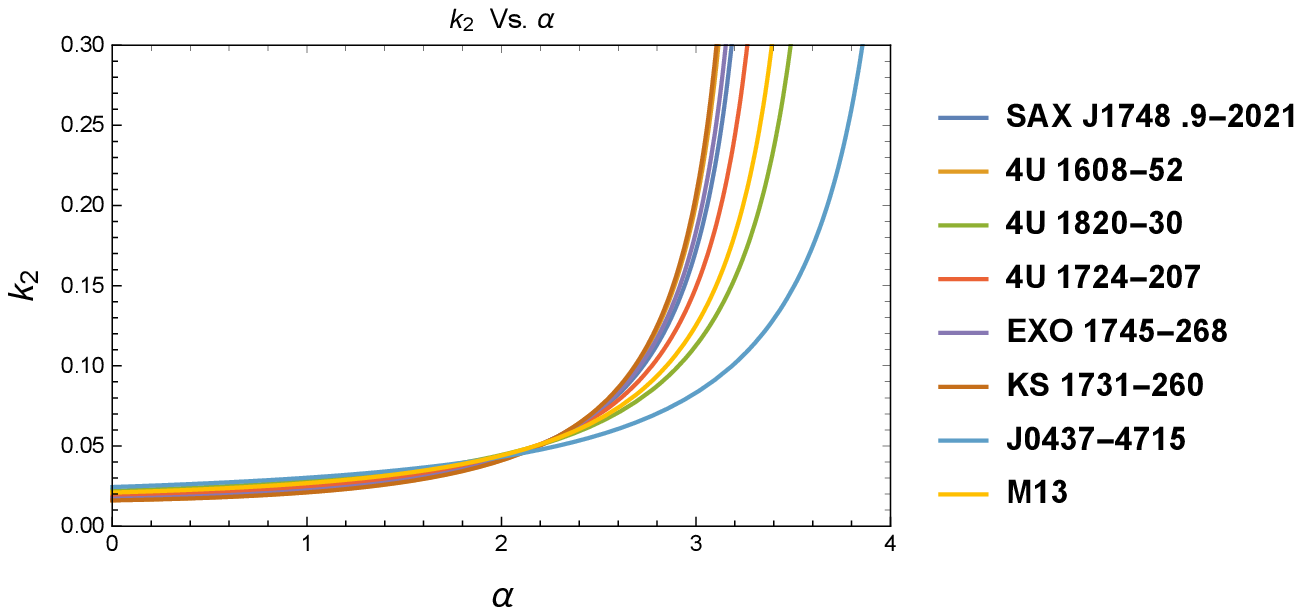}
	    \caption{$k_2$ is plotted against $\alpha$ for different compact objects.  Only physically allowed range of $\alpha$ is considered.}
	    \label{fig:b11}
	\end{figure}
	
	\begin{table*}
    \caption{ $y$ and $k_2$ for $\alpha = 0$ and the corresponding maximum value of $\alpha$are evaluated for different compact objects. The observational values of masses and radii of the compact stars have been considered from reference \cite{Roupas2020}.}
    \begin{tabular*}{\textwidth}{@{\extracolsep{\fill}}lrrrrrrrrl@{}}
    \hline
    Name & \multicolumn{1}{c}{ M($M_{\odot}$)} & \multicolumn{1}{c}{$R$ (km)} & \multicolumn{1}{c}{ $y$ at $\alpha = 0$} & \multicolumn{1}{c}{$k_2$ at $\alpha = 0$} & \multicolumn{1}{c}{Max. $\alpha$} & \multicolumn{1}{c}{$k_2$ at max. $\alpha$ }\\
    \hline \hline
     SAX J1748 .9-2021  &  $1.81^{+0.25}_{-0.37}$ & $11.7^{+1.7}_{-1.7}$ & 1.88994 & 0.0176463 & 3.006 & 0.174571  \\
      4U 1608-52   & $1.57^{+0.30}_{- 0.29}$ & $9.8^{+1.8}_{-1.8}$ & 1.93653 & 0.0162621 &  2.9341 & 0.167513  \\
      4U 1820-30 & $1.46^{+0.21}_{-0.21}$ & $11.1^{+1.8}_{-1.8}$ & 1.79387 & 0.0221002 &  3.3467 & 0.207726 \\
      4U 1724-207 & $1.81^{+0.25}_{-0.37}$ & $12.2^{+1.4}_{-1.4}$ & 1.85006 & 0.0190864 & 3.093 & 0.182933 \\
      EXO 1745-268  & $1.65^{+0.21}_{-0.31}$ & $10.5^{+1.6}_{-1.6}$ & 1.90908 & 0.0170474 &  2.973 & 0.171127 \\
      KS 1731-260 & $1.61^{+0.35}_{-0.37}$ & $10^{+2.2}_{-2.2}$ & 1.94436 & 0.0160516 & 2.924 & 0.166511  \\
     J0437-4715 & $1.44^{+0.07}_{-0.07}$ & $13.6^{+0.9}_{-0.8}$ & 1.78635 & 0.0243844 & 3.78 & 0.248741 \\
     M13 & $1.38^{+0.08}_{-0.23}$ & $9.95^{+0.24}_{-0.27}$ & 1.81066 & 0.0209619 & 3.2355 & 0.196986 \\
     \hline
     
    \end{tabular*} \label{bV}
\end{table*}

In the figure \ref{fig:b10}, variation of $y$ with respect to $\alpha$ is plotted for different compact stars. In this case, the range of $\alpha$ is assumed to be $0 	\leq \alpha < 4$. In figure \ref{fig:b11}, $k_2$ is plotted against $\alpha$. Physically acceptable range of $\alpha$ has been calculated numerically. Obviously, the maximum value of $\alpha$ is not the same for different class of compact stars. In table \ref{bV}, the maximum value of $\alpha$ is calculated for different neutron stars and the corresponding tidal love number $k_2$ is also shown. In figure \ref{fig:b11}, we note that the tidal love number increases monotonically with increasing $\alpha$ for stars having different compactness. In table \ref{bV}, $k_2$ is calculated for $\alpha = 0$ for different compact objects.

\section{\label{sec5} Discussion}
In this paper, we have presented a technique to measure the tidal love number of a compact object when subjected to an external tidal field. Possible role of anisotropy vis-a-vis matter distribution of the star on the tidal love number has been investigated. It remains to be seen whether such impact, if any, can be observationally realized. Moreover, effects of other factors such as EOS, electromagnetic field etc on the tidal love number needs further investigation and will be taken up elsewhere.

\section*{Acknowledgement}
SD and RS gratefully acknowledge support from the Inter-University Centre for Astronomy and Astrophysics (IUCAA), Pune, India, under its Visiting Research Associateship Programme.

\bibliographystyle{apsrev4-2}
	\bibliography{introrefv1}

\end{document}